\documentclass[11pt,a4paper]{article}

\usepackage{color}
\usepackage{amsfonts}
\usepackage{graphicx}

\title{\textbf{Kink dynamics in a system of two coupled scalar fields in two space-time dimensions}}

\author{A. Alonso Izquierdo$^{(a)}$
\\ {\normalsize {\it $^{(a)}$ Departamento de Matematica
Aplicada}, {\it Universidad de Salamanca, SPAIN}} }

\date{}

\begin{document}

\maketitle

\begin{abstract}
In this paper we examine the scattering processes among the members of a rich family of kinks which arise in a (1+1)-dimensional relativistic two scalar field theory. These kinks carry two different topological charges that determine the mutual interactions between the basic energy lumps (extended particles) described by these topological defects. Processes like topological charge exchange, kink-antikink bound state formation or kink repulsion emerge depending on the charges of the scattered particles. Two-bounce resonant windows have been found in the antikink-kink scattering processes, but not in the kink-antikink interactions.
\end{abstract}

\section{Introduction}

Over the last decades topological defects have played a key role in the understanding of new phenomena in a large number of disciplines of non-linear science. For this reason the search for these types of solutions in some PDEs receives much attention in both Mathematics and Physics, see \cite{Drazin1989,Rajaraman1982}. Among these equations the relativistic non-linear Klein-Gordon equation
\[
\frac{\partial^2 \phi_i}{\partial t^2} - \frac{\partial^2 \phi_i}{\partial x^2} = - \frac{\partial U}{\partial \phi_i} \hspace{0.5cm},\hspace{0.5cm} i=1,\dots,D
\]
is a prominent member, which has been extensively cited in the literature \cite{Manton2004,Vilenkin2000}. Here $D$ is the number of fields needed to describe a certain phenomenon. It is also referred to as the dimension of the internal space $(\phi_1,\dots,\phi_D)$. For instance, in Condensed Matter Physics $D$ is the number of order parameters in 1D materials, see \cite{Davydov1985}, where the dispersion relations in the different phonon branches are all of them relativistic. Here, the study of kinks in relativistic (1+1)-dimensional $D=2$ coupled scalar field theory is addressed.
For these systems an associated energy-momentum tensor can be found, which can be used to introduce a formal definition of kink: A kink is a non singular solution of the nonlinear coupled field equations of finite energy whose energy density $T_{00}$ is localized at any point in time \cite{Rajaraman1982}. This last characteristic portrays these solutions as energy lumps, of which each can be interpreted as \lq\lq extended particles\rq\rq{} of the model, which differentiates the kinks from plane wave packets (radiation), that delocalize its energy over time. The celebrated soliton and kink of the sine-Gordon and $\phi^4$ models \cite{Drazin1989,Rajaraman1982} are included in this context. These models encompass only one scalar field and provide theoretical support to explain, for instance, the appearance of superconductivity in type II materials \cite{Eschenfelder2012, Jona1993,Salje1993,Strukov2012}, electric charge fractionization in trans-polyacetylene (CH)$_x$ \cite{Golander1992}, the Josephson effect \cite{Davydov1985}, stabilization of interbrane spacings in brane world scenarios which resolve the problems of the cosmological constant and the large hierarchy between the scales of weak and gravitational forces \cite{DeWolfe2000:prd,Campos2002:prl}, among others. Promotion of these models to the quantum realm has led to studies on the one-loop corrections to the masses of these (1+1)-dimensional topological defects, see \cite{Dashen1974:prd, Dashen1975:prd, Alonso2002:npb, Alonso2012:tepjc}. A very interesting phenomenon, which deserves special attention, is the resonant kink-antikink interaction. In the seminal paper \cite{Campbell1983:pd}, Campbell, Schonfeld and Wingate thoroughly describe the dynamics of interacting kinks and antikinks in the $\phi^4$ model. There exists a critical velocity $v_c$ which characterizes the behavior of these scattering processes. For kink-antikink collisions where the initial speed is $v_0>v_c$ these single solutions collide, bounce back and escape. However, if $v_0<v_c$ they are compelled to collide a second time. Campbell and his collaborators discovered that there exist certain initial velocity windows where the kink and the antikink escape after the second impact, while at other windows they form a bound state. Other resonant windows, where the kink and the antikink escape after colliding $N\geq 3$ times, have also been found in this model. These authors also quantitatively explain this behavior by using collective coordinates for the kink solutions in this model. This effect can be explained by the resonant energy transfer mechanism, where an energy exchange between the kink translational mode and the internal vibrational mode takes place in each collision. At the first collision the internal vibrational eigenmode of the kink and antikink is excited, which decreases the kinetic energy. For particular initial velocities the energy of the vibrational mode is given back to the kinetic energy at subsequent collisions, which allows the kink and the antikink to escape in opposite directions. The application of the collective coordinate method to kink-antikink scattering in the $\phi^4$ model was initially addressed in \cite{Sugiyama1979:potp} and later corrected in \cite{Weigel2014:jop, Takyi2016:prd}. Another novel property unveiled in this work is the fractal structure followed by the separation velocity versus collision velocity graph \cite{Anninos1991:prd}. Goodman was able to provide a deep understanding of this feature in his illustrative papers \cite{Goodman2008:chaos,Goodman2005:sjoads,Goodman2007:prl}. He derives from the collective coordinate model for the resonant energy transfer mechanism a family of iterated maps which closely describe the chaotic behavior of the kink-antikink scattering. The previously described pattern is not specific to the $\phi^4$ model and has been found in many other situations such as in the kink-antikink interactions in the modified sine-Gordon model \cite{Peyrard1983:pd} and in the $\phi^6$ model \cite{Weigel2014:jop, Gani2014:prd}, the kink-impurity interactions in the sine-Gordon and $\phi^4$ models \cite{Fei1992:pra,Fei1992b:pra}, the soliton-defect interaction in the sine-Gordon model \cite{Goodman2002:pd, Goodman2004:pd, Javidan2006:jopa}, the interaction of kinks with local inhomogeneities \cite{Saadatmand2012:ps,Saadatmand2013:bjop} and the collision of vector solitons in the coupled nonlinear Schr\"odinger model \cite{Tan2001:pre,Yang2000:prl,Goodman2005:pre}. It is of note that the single kink in the $\phi^6$ model lacks internal vibrational modes \cite{Weigel2014:jop} and the resonant energy transfer mechanism is triggered by an internal vibrational mode of the combined kink-antikink configuration \cite{Dorey2011:prl}. In addition, the presence of many kink vibrational modes can lead to the suppression of bounce-windows in kink-antikink collisions \cite{Simas2016:johep} and the presence of quasiresonances \cite{Campbell1986:pd,Gani1999:pre}.

An enrichment of the previously mentioned models is reached by increasing the internal space dimension. This involves the presence of several scalar fields in the model, which can be coupled by a potential function $U(\phi_i)$. Rajaraman's quotation \textit{\lq\lq This already brings us to the stage where no general methods are available for obtaining all localized static solutions (kinks), given the field equations\rq\rq} \cite{Rajaraman1982} highlights the analytical difficulty in searching for kink solutions in this type of system. A huge amount of effort has been devoted to this issue in the last few decades, although most of this effort has been aimed at identifying static kink manifolds that exist in some field theory models. Here, the solutions depend only on the spatial coordinate and do not evolve in time. This static picture provides unchanging topological defects, where the forces acting on every point are balanced. It is assumed that an isolated basic \lq\lq particle\rq\rq{} described by a topological defect must be a member of this static kink manifold. This set may also include composite kinks, which consist of a distribution of basic kinks laid out in such a way that the total force exerted at every point vanishes. In other words, the static kink manifold provides us with a description of a set of solutions where the evolution of time is frozen.

In this paper we shall address the study of the kink dynamics in a two coupled scalar field theory model whose potential term is given by $U(\phi,\psi) = (4\phi^2+\psi^2-1)^2+4 \phi^2\psi^2$. This model arises as a special member of the one-parameter model family with $U(\phi,\psi) = (4\phi^2+2\sigma \psi^2-1)^2+16 \sigma^2 \phi^2\psi^2$ discussed by Bazeia and coworkers in the references \cite{Bazeia1995:pla, Bazeia1997:jopa}, where the authors identify a pair of topological kinks. Shifman and Voloshin showed that this family of systems can be found as the dimensional reduction of a generalized Wess-Zumino model with two chiral super-fields \cite{Shifman1998b:prd,Shifman1998:prd}. They found that the static kink manifold in this model comprises a one-parameter family of energy degenerate composite kinks. These solutions are formed by two basic kinks which belong to different topological sectors and whose centers are placed at distinct points. An explicit demonstration of the stability of some of these solutions is presented in \cite{Dias2007:ijompa}. In addition to this, Sakai and Sugisaka analytically explore the existence of bound states of wall-antiwall pairs \cite{Sakai2002:prd}. A supersymmetric version of this model compatible with local supersymmetry, where the kinks of the (1+1)-dimensional model promote to exact extended solutions of ${\cal N}=1$ (3+1)-dimensional supergravity was constructed in \cite{Eto2003:prd}. In this framework, the coupling of this scalar field theory model to gravity in (4+1)-dimensions in warped spacetime is considered by Bazeia in \cite{Bazeia2004:johep}. The formation of planar networks of topological defects is addressed in \cite{Bazeia2000:prd,Avelino2009:prd}. The breaking of the classical energy degeneracy for the static kink family by quantum-induced interactions has been studied in \cite{Alonso2004:npb, Alonso2014:jhep}. The distinctiveness of the case $\sigma=\frac{1}{2}$ in the previously mentioned model family was first noted in \cite{Alonso1998:jopa, Alonso2002b:prd} where it was shown that the analogue mechanical system derived from the static Klein-Gordon equation is Hamilton-Jacobi separable. This mechanical analogy underlies the fact that solving the static Klein-Gordon equation is tantamount to finding the solutions of
a Lagrangian dynamical system in which $x$ plays the role of time, the point-like particle position is determined by the fields $\phi_i$, and the potential energy of the particle is $-U$. This point allowed the authors to identify a second one-parameter family of energy degenerate static composite kinks for this particular member. These solutions are made of four basic particles (energy lumps). This means that an arrangement of four single kinks for which the mutual interactions are counterbalanced exists. A first attempt to study the kink dynamics for this model was accomplished in \cite{Alonso2002:prd} in the adiabatic approximation, where it is assumed that the kink motion is very slow. Under these circumstances the evolution of the particles or energy lumps can be studied as geodesics on the static kink moduli space \cite{Manton1982:plb, Manton2004, Tong2002:prd, Alonso2005:pd, Alonso2006:pd}. The goal of this paper is to study the interactions between the basic \lq\lq extended particles\rq\rq{} beyond the adiabatic approximation in the proposed model. It will be shown that in this two-scalar field theory model there exist eight single particles or energy lumps (described by four kinks and its corresponding antikinks) which carry two topological charges. Also, these particles will be scattered to uncover the nature of the interactions between them. Processes like topological charge exchange, kink-antikink bound state formation and kink repulsion emerge depending on the charges of the scattered particles.

The non-linearity of the evolution equations does not allow analytical tools to be employed to describe the behavior of the scattering solutions in detail. For this reason these processes will be studied by means of numerical simulations. Also, the modified algorithm described by Kassam and Trefethen in \cite{Kassam2005:sjosc} will be used, which has been designed to solve the numerical instabilities of the exponential time-differencing Runge-Kutta method introduced in \cite{Cox2002:jocp}. This explicit method is spectral in space and fourth order in time \cite{spectral}. A scattering process can be confined to occur in a bounded spatial interval, but the collisions between the energy lumps can originate radiation which travels at relativistic speeds and arrives to the spatial frontiers of the simulation in a short period of time. Therefore, the previous numerical method must be implemented in a spatial interval large enough to avoid the radiation to corrupt the results. In order to gain more control over the radiation evolution, the previous scheme will be complemented with the use of the energy conservative second-order finite difference Strauss-Vazquez algorithm \cite{Strauss1978:jocp} implemented with Mur boundary conditions \cite{Mur1981:emc}, which absorb the linear plane waves at the boundaries. The efficiency of this numerical scheme has been proved in \cite{Jimenez1990:amac}, and its global stability and convergence were established in \cite{PenYu1983:jas}. More general numerical schemes in this framework are studied in references \cite{Furihata2001:jocaam,Li1995:sjna,Fei1995:amac}.

The organization of this paper is as follows: in Section 2 the (1+1)-dimensional two coupled scalar field theory model, which is dealt with in this paper, will be introduced. The static kink manifold will be described where the basic topological defects of the model can be identified. These solutions involve just a single energy lump and can be interpreted as the \lq\lq basic extended particles\rq\rq{} of the system. These fundamental kinks are distinguished by the value of its topological charges and chirality, which endow them with different properties. The composite kink solutions included in the manifold are also explained in this section. In Section 3, the scattering processes between the basic particles will be analyzed. The strategy consists of colliding boosted kink solutions whose centers are initially placed far away. Four different scattering events arise, which depend on the topological charge values of the kinks involved in the collision. This allows the distinct interactions between the particles or lumps arising in the system to be described. In Section 4, the conclusions and final comments will be provided.

\section{The scalar field theory model and the static kink manifold}

We shall deal with a (1+1)-dimensional two-coupled scalar field theory model whose dynamics is governed by the action
\begin{equation}
S=\int d^2 x \left\{ \frac{1}{2} \partial_\mu \phi \, \partial^\mu \phi + \partial_\mu \psi \, \partial^\mu \psi -U(\phi,\psi) \right\} \hspace{0.3cm},
\label{action}
\end{equation}
where Einstein summation convention is assumed for $\mu=0,1$. Here $\phi:\mathbb{R}^{1,1} \rightarrow \mathbb{R}$ and $\psi:\mathbb{R}^{1,1} \rightarrow \mathbb{R}$ are dimensionless real fields and the Minkowski metric $g_{\mu\nu}$ in the two-dimensional spacetime is chosen in the form $g_{00}=-g_{11}=1$ and $g_{12}=g_{21}=0$. We shall denote the spacetime coordinates as $x^0\equiv t$ and $x^1 \equiv x$ from now on. The potential function $U(\phi,\psi)$ in (\ref{action}) is given by the non-negative expression
\begin{equation}
U(\phi,\psi) = (4\phi^2+\psi^2-1)^2+4 \phi^2\psi^2
\label{potential}
\end{equation}
for our model. The Euler-Lagrange equations derived from the action (\ref{action}) lead to the coupled nonlinear Klein-Gordon equations
\begin{eqnarray}
\frac{\partial^2 \phi}{\partial t^2} - \frac{\partial^2 \phi}{\partial x^2} = - \frac{\partial U}{\partial \phi} &=& - 16 \phi \left[4 \phi^2 + {\textstyle\frac{3}{2}} \psi^2 -1\right]  \hspace{0.2cm}, \label{euler01}\\
\frac{\partial^2 \psi}{\partial t^2} - \frac{\partial^2 \psi}{\partial x^2} = - \frac{\partial U}{\partial \psi} &=& - 4 \psi \left[6 \phi^2 + \psi^2 -1\right]\hspace{0.6cm}, \label{euler02}
\end{eqnarray}
which characterize the solutions of the system. The spatial and time translational symmetries, which arise in this type of scalar field theories, involve the conservation of the total energy $E[\phi,\psi]$ and momentum $P[\phi,\psi]$ defined as
\begin{eqnarray}
E[\phi,\psi]&=& \int dx \left[ \frac{1}{2} \Big( \frac{\partial \phi}{\partial t} \Big)^2 +  \frac{1}{2} \Big( \frac{\partial \psi}{\partial t} \Big)^2 +  \frac{1}{2} \Big( \frac{\partial \phi}{\partial x} \Big)^2 +  \frac{1}{2} \Big( \frac{\partial \psi}{\partial x} \Big)^2 + U(\phi,\psi) \right] \hspace{0.3cm}, \label{totalenergy} \\
P[\phi,\psi] &=& \int dx \left[ \frac{\partial \phi}{\partial x} \frac{\partial \phi}{\partial t} + \frac{\partial \psi}{\partial x} \frac{\partial \psi}{\partial t}\right] \hspace{0.3cm}. \label{totalmomentum}
\end{eqnarray}
In addition to the previous continuous symmetries, there exist discrete symmetries in our model. The action functional (\ref{action}) remains invariant by the symmetry group $\mathbb{G}=\mathbb{Z}_2\times \mathbb{Z}_2$ generated by the transformations $\pi_1 : (\phi,\psi) \mapsto (-\phi,\psi)$ and $\pi_2:(\phi,\psi) \mapsto (\phi,-\psi)$ in the internal space. The mirror reflection in the space coordinate $\pi_x:x\mapsto -x$ does also play an important role in the study of topological defects because it relates kink and antikink solutions.

The simplest stable solutions of equations (\ref{euler01}) and (\ref{euler02}) are the static homogenous solutions which correspond with the minima of the potential function $U(\phi,\psi)$. From (\ref{potential}), they constitute the set of zeroes of the potential function ${\cal M}=\{(\phi_0,\psi_0)\in \mathbb{R}^2: U(\phi_0,\psi_0)=0\}$:
\[
{\cal M}= \left\{ A_1=( {\textstyle\frac{1}{2}},0), A_2=(-{\textstyle\frac{1}{2}},0), B_1=(1,0) , B_2=(-1,0) \right\}\hspace{0.3cm},
\]
whose total energy is zero. Notice that $\pi_1(A_1)=A_2$ and $\pi_1(A_2)=A_1$ such that the transformation $\pi_1$ links the constant solutions $A_1$ and $A_2$. This allows us to define the vacuum orbit $\mathbf{A}=\{A_1,A_2\}$. Likewise, $\pi_2(B_1)=B_2$ and $\pi_2(B_2)=B_1$ and $\mathbf{B}=\{B_1,B_2\}$ is the second vacuum orbit in this model.

In general, the configuration space comprises the set of maps $\phi:\mathbb{R}^{1,1}\rightarrow \mathbb{R}$ and $\psi:\mathbb{R}^{1,1}\rightarrow \mathbb{R}$ whose total energy (\ref{totalenergy}) is finite, ${\cal C}=\{ \Phi(x,t)=(\phi(x,t),\psi(x,t))\in \mathbb{R}\times \mathbb{R} : E[\Phi(x,t)]<+\infty \}$. Owing to the total energy conservation law, the compliance of the previous condition in a single instant $t_0$ is sufficient for a map $\Phi(x,t)$ to belong to ${\cal C}$. Based on the previous definition, every member of ${\cal C}$ must satisfy the following asymptotic conditions
\begin{eqnarray}
&&\lim_{x\rightarrow \pm \infty} \frac{\partial \Phi(x,t)}{\partial t} =  \lim_{x\rightarrow \pm \infty} \frac{\partial \phi(x,t)}{\partial t} = \lim_{x\rightarrow \pm \infty} \frac{\partial \psi(x,t)}{\partial t} = 0 \hspace{0.3cm},\label{asymptotic01}\\ && \lim_{x\rightarrow \pm \infty} \frac{\partial \Phi(x,t)}{\partial x} =  \lim_{x\rightarrow \pm \infty} \frac{\partial \phi(x,t)}{\partial x} =  \lim_{x\rightarrow \pm \infty} \frac{\partial \psi(x,t)}{\partial x} = 0 \label{asymptotic02}\hspace{0.3cm}, \\[0.2cm] && \lim_{x\rightarrow \pm \infty} \Phi(x,t) = \lim_{x\rightarrow \pm \infty} \left(\phi(x,t),\psi(x,t)\right) \in {\cal M} \hspace{0.3cm}. \label{asymptotic03}
\end{eqnarray}
The following step will be to identify static kinks, time-independent finite-energy solutions of the field equations (\ref{euler01}) and (\ref{euler02}) whose energy density is localized. This type of solutions usually lives in non-zero topological sectors of the configuration space ${\cal C}$ and its spatial dependence asymptotically links two different elements of ${\cal M}$. This behavior allows two different topological charges to be introduced
\[
q_1= 2\cdot \left[\phi(+\infty,t_0)-\phi(-\infty,t_0)\right]  \hspace{0.5cm},\hspace{0.5cm} q_2=\psi(+\infty,t_0)-\psi(-\infty,t_0)\hspace{0.3cm},
\]
in our two scalar field theory model. These magnitudes are invariant because of the previous asymptotic conditions (\ref{asymptotic01})-(\ref{asymptotic03}). Solutions carrying non-zero topological charges are unable to evolve in time to zero-energy solutions (this would require infinite energy). In particular, the linear stability of a static solution $\Phi(x)$ is studied by means of the spectrum of the second order small fluctuation operator
\begin{eqnarray}
{\cal H}[\Phi(x)]&=&\left( \begin{array}{cc} -\frac{d^2}{dx^2} + V_{11}(x) & V_{12}(x) \\ V_{12}(x) & -\frac{d^2}{dx^2}+ V_{22}(x) \end{array} \right) = \label{hessiano}\\ &=& \left( \begin{array}{cc} -\frac{d^2}{dx^2} + \frac{\partial^2 U}{\partial \phi\, \partial \phi}[\Phi(x)] & \frac{\partial^2 U}{\partial \phi \,\partial \psi}[\Phi(x)] \\ \frac{\partial^2 U}{\partial \phi \,\partial \psi}[\Phi(x)] & -\frac{d^2}{dx^2}+ \frac{\partial^2 U}{\partial \psi\, \partial \psi} [\Phi(x)] \end{array} \right) \nonumber
\end{eqnarray}
The existence of negative eigenvalues in the spectrum of the operator (\ref{hessiano}) implies that the solution $\Phi(x)$ is unstable. Notice that (\ref{hessiano}) is a $2\times 2$ matrix partial differential operator making the identification of its spectrum an arduous task.

The identification of the static kink manifold in this model has been tackled in previous works from two points of view. The first procedure exploits the Hamilton-Jacobi separability of the analogue mechanical model derived from the static Klein-Gordon equations \cite{Alonso1998:jopa}. The second method makes use of the presence of two different superpotentials $W_{\rm I}=4\sqrt{2}(\frac{1}{3}\phi^3-\frac{1}{4}\phi+\frac{1}{4} \phi \psi^2)$ and $W_{\rm II}=\frac{1}{3}\sqrt{2} \sqrt{\phi^2+\psi^2} (4\phi^2+\psi^2-3)$, which lead to the same potential function (\ref{potential}) and reduce the problem (\ref{euler01}) and (\ref{euler02}) to first order differential equations, see \cite{Alonso2002b:prd}. The static kinks reported in \cite{Alonso1998:jopa,Alonso2002b:prd} are listed below and will be organized and displayed in a way that facilitates the intelligence of the kink dynamics, an issue that will be researched in the next section and represents the main goal of this work.

\vspace{0.2cm}

\noindent (1) We shall start this description by introducing the basic energy lumps or particles of the model, which are characterized by the eight static kinks
\begin{equation}
K_{\rm static}^{(q_1,q_2,\lambda)}(\overline{x})=\Big(  \frac{q_1}{4} \, \left[\lambda +\tanh ( \sqrt{2} \, \overline{x}) \right] , -\lambda \, q_2 \sqrt{{\frac{1}{2}}\, \left[1- \lambda \tanh [ \sqrt{2} \,\overline{x}] \right]} \Big) \label{basickink}
\end{equation}
where $q_i,\lambda=\pm 1$. In addition the compact notation $\overline{x}=x-x_0$ being $x_0\in \mathbb{R}$ the position of the kink center in the real line has been used. The parameter $x_0$ arises as an integration constant when solving the differential equations and proves the spatial translational invariance of the model. The parameters $q_i$ ($i=1,2$) in the expression (\ref{basickink}) are the topological charges associated with these kinks and take the values of $\pm 1$, see Figure 1. In any case these kinks asymptotically link elements in the vacuum orbit $\mathbf{A}=\{A_1, A_2\}$ (placed in the $\psi=0$ axis) with elements of the vacuum orbit $\mathbf{B}=\{B_1,B_2\}$ (located in the $\phi=0$ axis), see Figure 1. Indeed, the parameter $\lambda$ introduced in (\ref{basickink}) (which will be called chirality for reasons which will be clear later on) determines the sense of this connection, if the kink comes from points of the orbit $\mathbf{A}$ at $x=-\infty$ and arrives to a $\mathbf{B}$ type element at $x=\infty$ then $\lambda=-1$, while if the reverse sense takes place then $\lambda=1$. All the eight basic kinks (\ref{basickink}) share the same total energy,
\[
E[K_{\rm static}^{(q_1,q_2,\lambda)}(\overline{x})] = \frac{\sqrt{2}}{3}
\]
although the kink energy density distribution
\begin{equation}
\varepsilon[K_{\rm static}^{(q_1,q_2,\lambda)}(\overline{x})] = \frac{1}{8} \, {\rm sech}^4 [\sqrt{2} \, \overline{x}] \Big( 2+ \cosh[2\sqrt{2} \, \overline{x}] + \lambda \sinh[2\sqrt{2} \, \overline{x}]\Big) \label{basicdensity}
\end{equation}
distinguishes between kinks with different chirality $\lambda$. The energy density (\ref{basicdensity}) is localized around only one point in such a way that the $K_{\rm static}^{(q_1,q_2,\lambda)}(\overline{x})$ kink can be interpreted as a basic particle (a single energy lump in the space line), see Figure 1 (right). The exponential decay of the energy density is stronger when $x\rightarrow \infty$ than when $x\rightarrow -\infty$ for kinks with chirality $\lambda=-1$, as can observed in the energy density profile represented by a red dashed curve in Figure 1 (right). The opposite pattern is found for kinks with $\lambda=1$, see the blue solid curve in Figure 1 (right). From a physical point of view, this fact is evident because this type of kinks asymptotically connects vacua with different mass matrices. In other words $K_{\rm static}^{(q_1,q_2,1)} (\overline{x})$ kinks are more energetic at the right side than at the left side of the energy density peak. The converse behavior occurs for kinks with chirality $\lambda=-1$. In a static scenario (where forces are absent), all the kink solutions with the same chirality are undistinguishable.

\begin{figure}[h]
\centerline{\includegraphics[height=2.8cm]{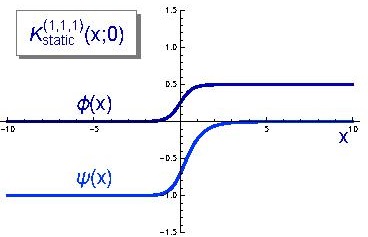} \hspace{0.2cm} \includegraphics[height=2.8cm]{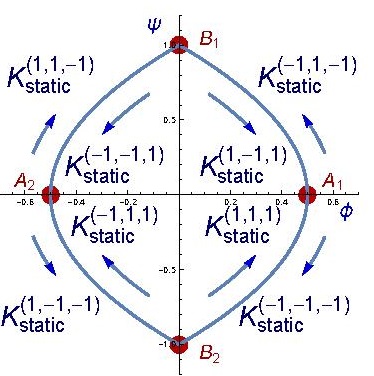}  \hspace{0.2cm} \includegraphics[height=2.8cm]{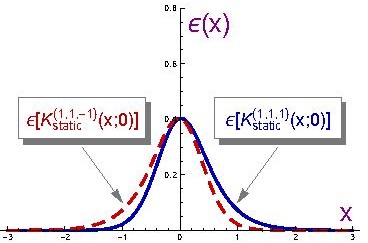}}
\caption{\small Scalar field components for the $K_{\rm static}^{(1,1,1)}(\overline{x})$ kink centered at the origin (left), the basic kink orbits connecting type \textbf{A} and \textbf{B} vacua (middle) and  energy density distribution for kinks with different chirality (right).}
\end{figure}

\noindent The action of the field reflection transformations $\pi_i$ ($i=1,2$) on the solutions (\ref{basickink}) reverses the sign of the topological charge $q_i$,
\begin{eqnarray*}
\pi_1[K_{\rm static}^{(q_1,q_2,\lambda)}(\overline{x})] &=& K_{\rm static}^{(-q_1,q_2,\lambda)}(x;x_0)\hspace{0.3cm}, \\
\pi_2[K_{\rm static}^{(q_1,q_2,\lambda)}(\overline{x})] &=& K_{\rm static}^{(q_1,-q_2,\lambda)}(x;x_0)\hspace{0.3cm}.
\end{eqnarray*}
The mirror or spatial reflection symmetry $\pi_x$ changes all the properties (both topological charges and chirality) of the basic kinks
\[
\pi_x[K_{\rm static}^{(q_1,q_2,\lambda)}(\overline{x})] = K_{\rm static}^{(-q_1,-q_2,-\lambda)}(\overline{x})\hspace{0.3cm},
\]
although the kink orbit remains unchanged (which is traced in reverse order).

\noindent By convention we shall refer to solutions (\ref{basickink}) with negative chirality $K_{\rm static}^{(q_1,q_2,-1)}(\overline{x})$ as \textit{kinks} and those with positive chirality $K_{\rm static}^{(q_1,q_2,1)}(\overline{x})$ as \textit{antikinks} where $q_i=\pm 1$. In addition, the term \textit{kink and antikink of the same type} will be used for those solutions which share the same orbit; that is, for the couple of kinks $K_{\rm static}^{(q_1,q_2,-1)}(\overline{x})$ and $K_{\rm static}^{(-q_1,-q_2,1)}(\overline{x})$, which are related by the transformation $\pi_x$, otherwise it will be said that the \textit{kinks and antikinks are of different types}. As shown later on, these distinctions simplify the language used in the scattering study.

The kink fluctuation operator ${\cal H}[K_{\rm static}^{(q_1,q_2,\lambda)}(\overline{x})]$ follows the form (\ref{hessiano}) where the potential wells $V_{ij}(x)$, $i,j=1,2$ are given by
\begin{eqnarray*}
V_{11}(x)&=& 4 [2+3\tanh(\sqrt{2}x)(\lambda+\tanh(\sqrt{2} x))] \\
V_{12}(x)&=& -6\sqrt{2} q_1 q_2 (1+\lambda \tanh(\sqrt{2} x))\sqrt{1-\lambda \tanh(\sqrt{2}x)} \\
V_{22}(x)&=& \frac{1}{2} [7+3\tanh(\sqrt{2}x)(-2\lambda+\tanh(\sqrt{2} x))]
\end{eqnarray*}
which have been depicted in the Figure 2 for the values $\lambda=q_1 q_2=-1$. The zero mode $\frac{\partial}{\partial x} K_{\rm static}^{(q_1,q_2,\lambda)}(\overline{x})$ is part of the spectrum as a consequence of the translational symmetry of the model. Numerical studies point out that the rest of the spectrum is a continuous spectrum on the threshold value $\omega^2=2$. No internal vibrational eigenmodes for the single kink appear in this model.

\begin{figure}[h]
\centerline{\includegraphics[height=2.8cm]{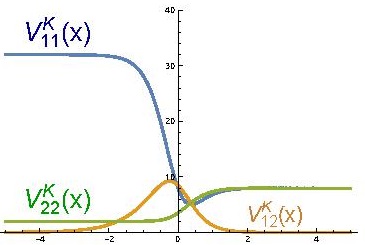}}
\caption{\small Potential well components $V_{ij}(x)$, $i,j=1,2$, of the kink fluctuation operator ${\cal H}[K_{\rm static}^{(q_1,q_2,\lambda)}(\overline{x})]$.}
\end{figure}

\vspace{0.2cm}

\noindent The static picture of the kink manifold of the model does not finish here. Furthermore, together with the basic kinks, there is a pair of two-parameter families of static composite kinks which are described below.

\vspace{0.2cm}

\noindent (2) The first of the previously mentioned families, which links the points $A_1$ and $A_2$, is determined by the expression
\begin{equation}
\overline{K}_{\rm static}^{(q_1,0)}(\overline{x},b)= \Big( \frac{q_1}{4} \frac{\sinh (2\sqrt{2}\,\overline{x})}{\cosh (2\sqrt{2}\,\overline{x})+b^2}, \frac{b}{[b^2+ \cosh (2\sqrt{2} \,\overline{x})]^\frac{1}{2}} \Big) \hspace{0.2cm}, \hspace{0.1cm} (q_1=\pm 2,\,\,b\in \mathbb{R}) \label{kink2}\hspace{0.1cm}.
\end{equation}
The magnitude $q_1$ inserted in (\ref{kink2}) is the first topological charge associated with these kinks, which ranges the values $2$ and $-2$. The second topological charge vanishes for these solutions. Every member of the kink family (\ref{kink2}) can be interpreted as the concatenation of two basic static kinks following the form
\begin{equation}
K_{\rm static}^{(q_1,q_2,-1)}(x-x_1) \cup K_{\rm static}^{(q_1,-q_2,1)}(x-x_2) \hspace{1cm} \mbox{with} \hspace{1cm} x_1\leq x_2\hspace{0.3cm},
\label{configuration01}
\end{equation}
that is, a kink with topological charges $(q_1,q_2)$ followed by an antikink with the charges $(q_1,-q_2)$ in the spatial coordinate $x$. Notice that the involved kink and antikink are of different types because they share the same charge $q_1$. The parameter $b$ in (\ref{kink2}) measures the distance between these two basic lumps. All of these features are illustrated in Figure 3. Take note that the energy density distribution for the solution (\ref{kink2}) with $b=5$ displayed in Figure 3 (middle) consists of two single energy lumps.

\begin{figure}[h]
\centerline{\includegraphics[height=2.8cm]{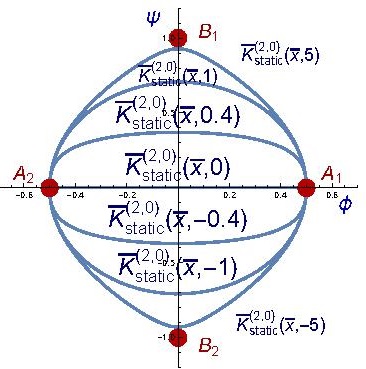} \hspace{0.2cm} \includegraphics[height=2.8cm]{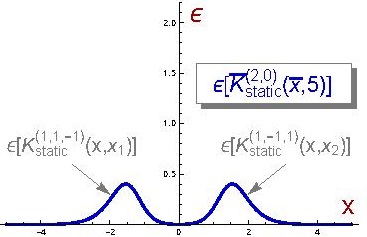} \hspace{0.2cm} \includegraphics[height=2.8cm]{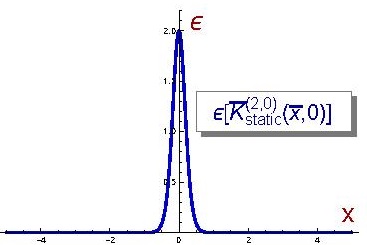}}
\caption{\small Orbits of the kink family $\overline{K}_{\rm static}^{(q_1,0)}(\overline{x},b)$ for several values of the parameter $b$ (left) and energy density distributions for the $\overline{K}_{\rm static}^{(q_1,0)}(\overline{x},5)$-kink (middle) and the $\overline{K}_{\rm static}^{(q_1,0)}(\overline{x},0)$-kink (right).}
\end{figure}

\noindent In particular, the solution $\overline{K}_{\rm static}^{(q_1,0)}(\overline{x},0)$, obtained from (\ref{kink2}) with $b=0$,
\begin{equation}
\overline{K}_{\rm static}^{(q_1,0)}(\overline{x},0)= \left({\textstyle\frac{q_1}{4}} \tanh (2\sqrt{2}\,\overline{x}),0 \right)
\label{kink20}
\end{equation}
describes a $(q_1,q_2)$-kink and a $(q_1,-q_2)$-antikink whose centers are placed at the same point. The energy density of this particular solution is represented in Figure 3 (right). The simplicity of the expression (\ref{kink20}) allows the linear stability of the solution $\overline{K}_{\rm static}^{(q_1,0)}(\overline{x},0)$ to be studied. In this case, the small kink fluctuation operator
\begin{equation}
{\cal H}[\overline{K}_{\rm static}^{(q_1,0)}(\overline{x},0)]= \left( \begin{array}{cc} -\frac{d^2}{dx^2} + 32 -48 \, {\rm sech}^2 2\sqrt{2}\,\overline{x} & 0 \\ 0 & -\frac{d^2}{dx^2} + 2 -6 \, {\rm sech}^2 2\sqrt{2} \, \overline{x}  \end{array} \right)
\label{fluctuation01}
\end{equation}
is a second order differential diagonal matrix operator. The discrete spectrum of (\ref{fluctuation01}) comprises two zero modes $\xi_0^T (x)=({\rm sech}^2 2 \sqrt{2} \overline{x},0)$ and $\xi_0'{}^T=(0,{\rm sech}^\frac{1}{2} 2\sqrt{2} \overline{x})$, which are in turn the ground states. Indeed, the presence of two zero modes occurs not just for the solution (\ref{kink20}) but for every member of (\ref{kink2}). This result underlies the fact that the kink family (\ref{kink2}) depends on two real parameters $x_0$ and $b$. By changing the value of one of these parameters a shift is made from solutions to solutions of the model. If the change is infinitesimal, the difference between the new and the original solutions becomes a zero mode. Therefore $\frac{\partial}{\partial x_0} \overline{K}_{\rm static}^{(q_1,0)}(\overline{x},b) \equiv \frac{\partial}{\partial x} \overline{K}_{\rm static}^{(q_1,0)}(\overline{x},b)$ and $\frac{\partial}{\partial b} \overline{K}_{\rm static}^{(q_1,0)}(\overline{x},b)$ correspond to the analytical expression of these eigenmodes. The first of these modes arises because of the spatial translational symmetry, and infinitesimally changes the kink center $x_0$ without changing the kink orbit. The second one deforms the orbit of the original solution, giving rise to a kink family member infinitesimally close in the $b$-parameter space. The final result of this change is that the basic kinks (which constitute this solution) separate infinitesimally. The excited state $\xi_1^T=({\rm sech}\,2\sqrt{2}x\tanh 2\sqrt{2}x , 0)$ with eigenvalue $\omega_1^2=24$ completes the discrete spectrum of (\ref{fluctuation01}). The excitation of this mode induces an internal vibration, the energy lump contracts (making its peak higher) and then stretches (lowing the energy peak) in a periodic sequence. The lack of negative eigenvalues in the spectrum of the fluctuation operator (\ref{fluctuation01}) implies that the particular kink (\ref{kink20}) is stable. In references \cite{Alonso1998:jopa,Alonso2002b:prd} it has been proven by applying the Morse theory on the kink orbit manifold that the previous conclusion is valid for every member of the family (\ref{kink2}).

\noindent (3) There exists another two-parameter family of static composite kink solutions, determined by the expression
{\begin{equation}
\overline{\overline{K}}_{\rm static}^{(0,q_2)}(\overline{x},c)=\Big( \frac{\sinh 2\sqrt{2} c \, \sinh 2\sqrt{2} \overline{x}}{\cosh^2 2\sqrt{2} \overline{x} + 2 \cosh 2\sqrt{2} c \cosh 2\sqrt{2} \overline{x}+1},\frac{q_2}{2} \frac{\sinh 2\sqrt{2} \overline{x}}{[\cosh^2 2\sqrt{2} \overline{x} + 2 \cosh 2\sqrt{2} c \cosh 2\sqrt{2} \overline{x}+1]^\frac{1}{2}} \Big)
\label{kink4}
\end{equation}}
where the parameter $c\in \mathbb{R}$ and $q_2$ is the second topological charge associated with these solutions whose possible values are $\pm 2$. These kinks live in the topological sector which joins the members $B_1$ and $B_2$ of the vacuum orbit $\mathbf{B}$. As a result the first topological charge $q_1$ vanishes. In Figure 4, the energy density of two members of this family is displayed. It can be observed that the solutions (\ref{kink4}) consist of four basic lumps following the antikink-kink-antikink-kink arrangement
\begin{equation}
K^{(q_1,q_2,1)}_{\rm static}(\overline{x}-x_1) \cup K_{\rm static}^{(-q_1,\overline{q}_2,-1)}(\overline{x}) \cup K_{\rm static}^{(-q_1,-\overline{q}_2,1)}(\overline{x}) \cup K_{\rm static}^{(q_1,q_2,-1)}(\overline{x}+x_1)\, ,
\label{configuration02}
\end{equation}
where the kink and antikink in the middle of this sequence are exactly overlapped (giving rise to a $\overline{K}_{\rm static}^{(q_1,0)}(\overline{x},0)$ configuration) and the rest ones are equidistant from this central lump. The distance between these constituents is set by the value of the family parameter $c$, see Figure 4.

\begin{figure}[h]
\centerline{\includegraphics[height=2.8cm]{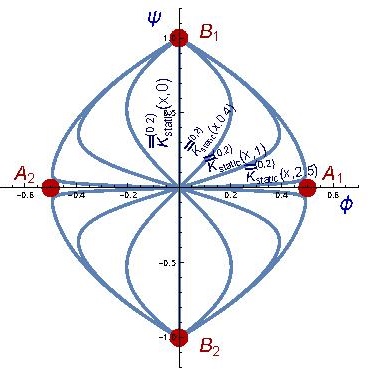} \hspace{0.2cm} \includegraphics[height=2.8cm]{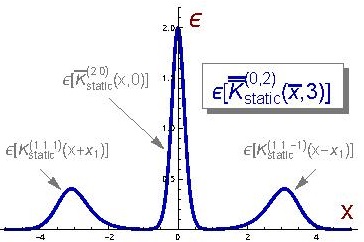}  \hspace{0.2cm} \includegraphics[height=2.8cm]{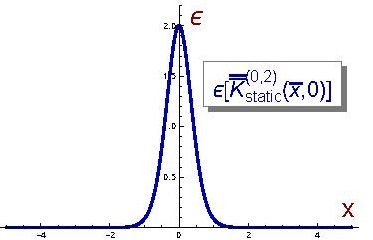}}
\caption{\small Orbits of the kink family $\overline{\overline{K}}_{\rm static}^{(0,q_2)}(\overline{x},c)$ for several values of the parameter $c$ (left) and energy density distributions for the $\overline{\overline{K}}_{\rm static}^{(0,q_2)}(\overline{x},3)$-kink (middle) and the $\overline{\overline{K}}_{\rm static}^{(0,q_2)}(\overline{x},0)$-kink (right).}
\end{figure}

\noindent For the special value $c=0$ the four basic kinks are located at the same point, and the solution (\ref{kink4}) reduces to the expression
\begin{equation}
\overline{\overline{K}}_{\rm static}^{(0,q_2)}(\overline{x},0) =\left(0, {\textstyle\frac{q_2}{4}} \tanh \sqrt{2} \overline{x} \right)
\label{kink40}
\end{equation}
whose density energy is displayed in Figure 4 (right). The evolution of the $\overline{\overline{K}}_{\rm static}^{(0,q_2)}(x,0)$-kink fluctuations is coded in the spectrum of the second order differential matrix operator
\begin{equation}
{\cal H}[\overline{\overline{K}}_{\rm static}^{(0,q_2)}(x,0)]= \left( \begin{array}{cc} -\frac{d^2}{dx^2} + 8 -24 \, {\rm sech}^2 \sqrt{2}\, \overline{x} & 0 \\ 0 & -\frac{d^2}{dx^2} + 8 -12 \, {\rm sech}^2 \sqrt{2} \, \overline{x}  \end{array} \right)\hspace{0.3cm}.
\label{fluctuation02}
\end{equation}
The discrete spectrum of (\ref{fluctuation02}) begins with a negative eigenvalue $\omega_0^2=-10$, whose eigenfunction is given by $\eta_0^T=({\rm sech}^3 \sqrt{2} \overline{x},0)$. In addition, two degenerate zero modes $\eta_1^T = ({\rm sech}^2 \sqrt{2} \overline{x} \tanh \sqrt{2} \overline{x}, 0)$ and $\eta_1'{}^T=(0,{\rm sech}^2 \sqrt{2}\overline{x})$ with zero eigenvalue $\omega_1^2=0$ are involved. The excitation of the longitudinal and transversal modes $\eta_1'{}^T$ and $\eta_1^T$ infinitesimally moves the family parameters $x_0$ and $c$. The action of $\eta_1'{}^T$ is a simple translation of the kink solution, while the action of $\eta_2'{}^T$ consists of separating an antikink and a kink from the original energy lump formed by four merged basic particles. An eigenvalue $\omega_2^2=6$ is also included in the discrete spectrum of (\ref{fluctuation02}) with the degenerate eigenfunctions $\eta_2^T=({\rm sech}\sqrt{2}\overline{x} (4-5\,{\rm sech}^2 \sqrt{2} \overline{x}),0)$ and $\eta_2'{}^T=(0,{\rm sech}\,\sqrt{2}\overline{x} \tanh \sqrt{2} \overline{x})$. These correspond to internal vibrational modes of the kink. Finally a continuous spectrum emerges on the threshold value $\omega^2=8$. The presence of the negative eigenvalue $\omega_0^2$ implies that the kink (\ref{kink40}) is unstable. This seems also to be supported by the fact that for this solution a kink $K_{\rm static}^{(q_1,q_2,-1)}(\overline{x})$ and its own antikink $K_{\rm static}^{(-q_1,-q_2,1)}(\overline{x})$ are placed at the same point. It should be noted that the evolution of the kink (\ref{kink40}) excited by the mode $\eta_0^T$ has not been described here because now the fluctuation grows indefinitely and the original kink changes into a completely different configuration. The application of the Morse theory indicates that all of the members of the family (\ref{kink4}) are unstable due to the presence of a conjugate point in the kink orbit space, see \cite{Alonso1998:jopa,Alonso2002b:prd}.

In summary, the study of the static kink manifold in this model reveals that there are four basic particles of the model characterized by the value of the pair of topological charges $(q_1,q_2)$ with $q_i=\pm 1$ and the chirality $\lambda=-1$. The corresponding antiparticles emerge by changing the sign of the topological charges and chirality. The static picture of the solutions described in this section points out that two basic particles of the type $K_{\rm static}^{(q_1,q_2,-1)}(\overline{x})$ and $K_{\rm static}^{(q_1,-q_2,1)}(\overline{x})$ can be placed at any location in the real line, giving rise to the family (\ref{kink2}). If these lumps stand still the configuration remains unchanged over time because all of the forces are balanced. Additionally, there is another configuration with an analogous behavior described by the family (\ref{kink4}). Here the composite kink $\overline{K}_{\rm static}^{(q_1,0)}(\overline{x};x_0,0)$ is surrounded by an antikink and a kink carrying topological charge $-q_1$ and equidistant to this central lump. In this case, all the local forces are again counteracted; however, this is an unstable configuration which is spoilt if a little perturbation is introduced on this arrangement.

\section{Kink dynamics: a study of two basic kink scattering}

In this section the kink dynamics for the two-scalar field theory model introduced in Section 2 is investigated, such that complete intelligence of the effect of the nonlinearity on the kink evolution can be achieved in this model. In Section 2, the existence of four basic particles and the corresponding antiparticles are unveiled, which are distinguished by chirality, even in the static framework. In this section the aim is to uncover kink interactions by studying the scattering of the basic lumps or particles of the model. In particular, the results displayed in Section 2 present some of the following questions:

\begin{enumerate}
\item What is the evolution of the static configuration (\ref{configuration01}) if the basic lumps are pushed? In other words, how does a kink with charges $(q_1,q_2)$ and an antikink with charges $(q_1,-q_2)$ evolve when they are scattered each other?

\item Similarly, how does an antikink with charge $(q_1,q_2)$ and kink with charge $(-q_1,q_2)$ behave when they are propelled against each other? There are no static kink configurations with this arrangement, so the permanent presence of forces in this process is presumed.

\item The fate of the collision between a kink and its own antikink is also unknown. In some models collision speeds less than a critical velocity produce a bound state, where kink and antikink are trapped in an oscillatory movement, while in other models this event ends in mutual annihilation. In addition, the resonant energy transfer mechanism can appear, giving rise to resonant initial velocity windows, which allow the kink and antikink ro escape after a finite number of collisions.
\end{enumerate}

\noindent All of these questions will be addressed in this section by employing a numerical analysis. However, before tackling this task this subsection will end by introducing some comments regarding the boosted basic kinks. In a relativistic system, the static kinks (\ref{basickink}) can be endowed with a time dependence by introducing a Lorentz boost
\begin{equation}
K^{(q_1,q_2,\lambda)}(\overline{x},t;v_0) = K_{\rm static}^{(q_1,q_2,\lambda)}\Big(\frac{\overline{x}-v_0t}{\sqrt{1-v_0^2}}\Big)\hspace{0.3cm},
\label{dynkink}
\end{equation}
such that the lumps move with constant velocity $v_0$. The total energy is now increased by the Lorentz factor
\[
E[K^{(q_1,q_2,\lambda)}(\overline{x},t;v_0)]= \frac{\sqrt{2}}{3 \sqrt{1-v_0^2}}\hspace{0.3cm}.
\]
Also, when the speed $v_0$ is increased, length contraction implies that the energy lumps are concentrated in a smaller region. This behavior is illustrated in Figure 5 for several velocities $v_0$.

\vspace{0.2cm}

\begin{figure}[h]
\centerline{\includegraphics[height=3cm]{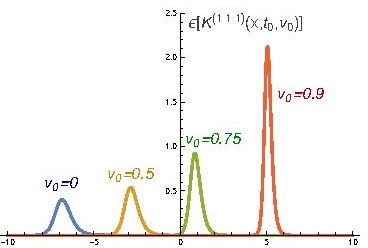}}
\caption{\small Energy density for several boosted basic kinks. Take note that the higher the velocity is, the thinner and taller the kink energy density is.}
\end{figure}

In the following sections the results obtained by the numerical simulations, carried out with the purpose of describing the interactions between the basic particles in this model, will be discussed. The initial configuration must consist of a concatenation of alternating basic kinks and antikinks because of the topological constraints. The trajectory for this type of profiles asymptotically begins in the vacuum orbit of a given type, \textbf{A} or \textbf{B}, then travels towards the vacuum orbit of the other type, \textbf{B} or \textbf{A}, to later return to the vacuum orbit of the first type, see Figure 1(middle). The scattering processes related by the reflection transformations $\pi_i$, $i=1,2$ and $\pi_x$ are equivalent. The evolution of the particles in a given event can be derived from an equivalent one. As a result, the present study will be restricted to non-equivalent processes. Furthermore, the analysis of the interactions between two of the basic kinks and antikinks, which can be considered as the fundamental events, will be discussed. These interactions are distinguished by the kink-antikink or antikink-kink arrangements, and by the relation between its types. We recall that a kink and an antikink are said to be of the same type if they share the same orbit, otherwise they are considered as different types.

\subsection{The $K^{(q_1,q_2,-1)}$-$K^{(q_1,-q_2,1)}$ scattering process: an exchange of the second topological charge}

Here we shall deal with the scattering of the two basic lumps which comprise the static composite kink family (\ref{kink2}). In this case, a basic kink is placed on the left of a basic antikink in the $x$-axis. These are chosen to be of different type. In Section 2 it is shown that if these basic lumps stand motionless, the dynamics leaves this situation unchanged. Following on, the evolution of these kinks when they are obliged to collide with each other at speed $v_0$ is analyzed. If the mass center is fixed at the spatial coordinate origin the initial configuration can be represented by the kink-antikink concatenation $K^{(q_1,q_2,-1)}(x+x_1,t;v_0) \cup K^{(q_1,-q_2,1)}(x-x_1,t;-v_0)$ where $K^{(q_1,q_2,\pm 1)}(\overline{x},t;v_0)$ is defined in (\ref{dynkink}) and $x_1$ is large enough to generate a continuous profile. The evolution of these basic lumps is displayed in Figure 6, where velocity $v_0$ has been chosen as $v_0=0.2$. It can be observed that the kink with charge $(q_1,q_2)$ and the antikink with charge $(q_1,-q_2)$ approach each other, coalesce, giving rise to an energy density sharper than the sum of the individual lump energy densities due to nonlinear interactions, finishing with the lumps bouncing back and exchanging the second topological charge $q_2$. This scattering process is represented by the relation
\[
K^{(q_1,q_2,-1)}(v_0) \cup K^{(q_1,-q_2,1)}(-v_0) \rightarrow K^{(q_1,-q_2,-1)}(-v_0) \cup K^{(q_1,q_2,1)} (v_0)
\]
where the velocities of each lump are indicated. The final outcome consists of a kink with charge $(q_1,-q_2)$ that travels to the left and an antikink with charge $(q_1,q_2)$ that moves to the right. This phenomenon is quite an elastic event, because it takes place with a negligible emission of radiation.

\begin{figure}[h]
\centerline{\includegraphics[height=4cm]{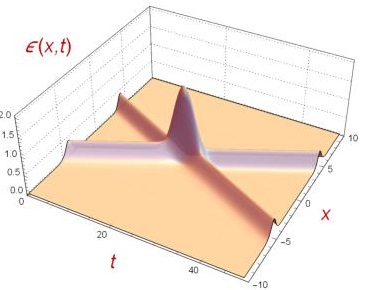} }
\caption{\small Energy density representation in the $K^{(q_1,q_2,-1)}$-$K^{(q_1,-q_2,1)}$ scattering process for the initial velocity $v_0=0.2$.}
\end{figure}

\noindent In reference \cite{Alonso2002:prd} the above kink-antikink scattering process has been addressed under the adiabatic approximation where very slow motion is assumed for the lumps following the Manton's scheme \cite{Manton1982:plb}. In this context it is assumed that the kink family parameters $x_0$ and $b$, which arise in the expression (\ref{kink2}), depend on time. This approach postulates that under the assumption of slowness of the process, a composite kink (\ref{kink2}) evolves into a new configuration characterized by the same expression (\ref{kink2}) although with different parameter values. If we plug this generalized form of (\ref{kink2}) into the action functional (\ref{action}) ordinary differential equations for the variables $x_0(t)$ and $b(t)$ are obtained. The kink dynamics is now described by geodesics on the two dimensional static kink moduli space $(x_0,b)$. In other words, the dynamics is ruled by the excitation of the two zero modes associated with these composite kinks (\ref{kink2}). The results found in this subsection endorse the description of the scattering introduced in \cite{Alonso2002:prd}. Indeed, they conclude that the validity of this analysis is applicable beyond the adiabatic approximation; that is, the description of the previous kink scattering is valid for an extensive range of collision velocities $v_0$. Simulations with $v_0=0.9$ show that the previously mentioned pattern is maintained; however, a very small amount of kinetic energy is now converted to radiation.

\subsection{The $K^{(q_1,q_2,1)}$-$K^{(-q_1,q_2,-1)}$ scattering process: repulsive forces in action}

In this subsection the scattering between an antikink with charge $(q_1,q_2)$ and a kink with charge $(-q_1,q_2)$ is analyzed. Now the basic lumps obliged to collide have the opposite first topological charge but the second topological charge is the same. The initial configuration consists of an antikink placed on the left of a different type of kink in the $x$-coordinate. We fix the mass center at the origin of the spatial axis such that the speed of each kink lump is $v_0$ in this reference frame. The initial arrangement can be represented by means of the antikink-kink concatenation $K^{(q_1,q_2,1)}(x+x_1,t;v_0) \cup K^{(-q_1,q_2,-1)}(x-x_1,t;-v_0)$. In Figure 7 (left) the evolution of this profile, where $v_0=0.2$, is displayed. The particles approach each other but when they are close enough the lumps repel each other, avoiding collision. This involves the presence of repulsive forces between antikinks with charge $(q_1,q_2)$ and kinks with charge $(-q_1,q_2)$. The resulting configuration is similar to the original one although now the lumps move away from each other. Therefore this process can be represented as
\[
K^{(q_1,q_2,1)} (v_0) \cup K^{(-q_1,q_2,-1)}(-v_0) \rightarrow K^{(q_1,q_2,1)} (-v_0) \cup K^{(-q_1,q_2,-1)} (v_0)\hspace{0.3cm}.
\]
The previous pattern is general for any initial velocity $v_0$. In Figure 8 (left) the final velocity of the lumps after the scattering process is represented with respect to $v_0$. The dashed line in  Figure 8 (left) characterizes the result if the process is elastic. Notice that these curves are indistinguishable for initial speeds $v_0$ up to 0.7. In these cases the radiation emission is negligible and the kink scattering is practically elastic.

\begin{figure}[h]
\centerline{\includegraphics[height=4cm]{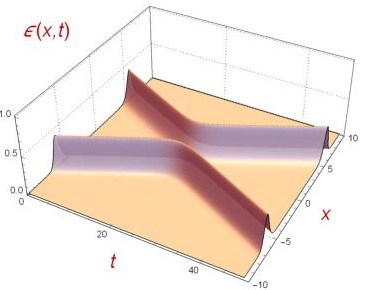} \hspace{1.5cm} \includegraphics[height=4cm]{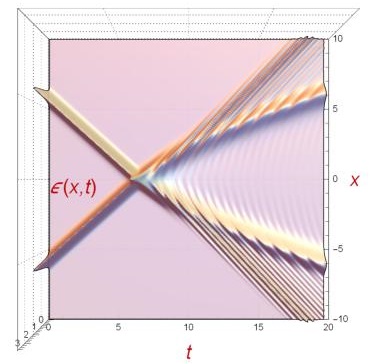}}
\caption{\small Energy density representation in the $K^{(q_1,q_2,1)}$-$K^{(-q_1,q_2,-1)}$ scattering process for the initial velocities $v_0=0.2$ (left) and $v_0=0.9$ (right).}
\end{figure}

\noindent However, for higher initial velocities radiation phenomena are appreciable. The collision between the antikink and the kink is now so violent that a part of the kinetic energy of these extended particles is emitted in form of radiation, which decelerates the lumps. In Figure 7 (right), this type of events is illustrated for the case $v_0=0.9$ using a top view in order to enhance the visualization of the process. It can be observed that after impact each basic particle emits radiation in both directions, where part of this radiation is trapped between the lumps because a large amount of it is reflected upon reaching the kink cores. The rest of the radiation advances towards the simulation frontiers. In Figure 8 (right), the minimum distance $d_{\rm min}$ between the lumps in the scattering process is plotted as a function of the initial velocity $v_0$. Both of the graphs in Figure 8 have been generated by means of a discrete number of points with the initial velocity step $\Delta v_0=0.01$.

\begin{figure}[h]
\centerline{\includegraphics[height=2.5cm]{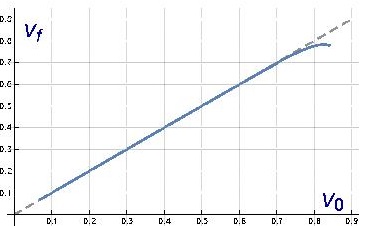} \hspace{1.5cm} \includegraphics[height=2.5cm]{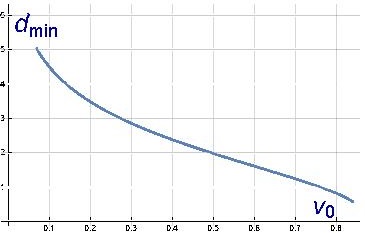}}
\caption{\small Graphic representation of the final velocity (left) and the minimal distance (right) of the lumps in the $K^{(q_1,q_2,1)}$-$K^{(-q_1,q_2,-1)}$ scattering as functions of the initial velocity $v_0$.}
\end{figure}

\subsection{The $K^{(q_1,q_2,-1)}$-$K^{(-q_1,-q_2,1)}$ scattering process}

In this subsection the scattering between a kink and an antikink of the same type is studied. In this case, the initial kink configuration is given by a kink with charge $(q_1,q_2)$ (this solution asymptotically starts at the point $A_i$ and arrives to the point $B_j$ following a given orbit) which is placed to the left of the antikink with charge $(-q_1,-q_2)$ (this solution returns from $B_j$ to the initial point $A_i$ retracing the previous kink trajectory in converse sense). Thus, this situation is dealing with the collision of a kink and its own antikink, which are pushed together with a velocity $v_0$. As usual we fix the mass center at the origin of the $x$-axis. The initial configuration is represented by the kink-antikink concatenation $K^{(q_1,q_2,-1)}(x+x_1,t;v_0) \cup K^{(-q_1,-q_2,1)}(x-x_1,t;-v_0)$. Here we can distinguish two different types of scattering events, which are separated by the critical velocity
\[
v_c \approx 0.29703
\]
in the initial velocity space. We find that:
\begin{itemize}
\item If $v_0<v_c$ the involved particles are trapped in a bound state (bion) where the kink and the antikink are forced to approach and bounce back over and over again. In Figure 9 (left) this event has been depicted for $v_0=0.2$. In this process a part of the total energy is converted into radiation (observe the small ripples in the Figure). This process is represented by means of the relation
\[
K^{(q_1,q_2,-1)}(v_0) \cup K^{(-q_1,-q_2,1)}(-v_0) \rightarrow K^{(q_1,q_2,-1)} \uplus K^{(-q_1,-q_2,1)}  + \mbox{radiation}
\]
where the use of the symbol $\uplus$ emphasizes the formation of the kink-antikink bion.

\begin{figure}[h]
\centerline{\includegraphics[height=4cm]{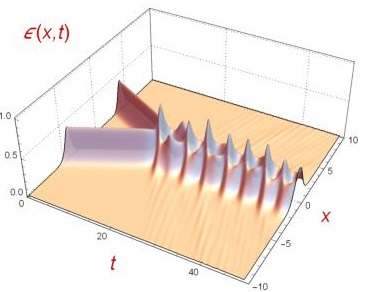} \hspace{1.cm} \includegraphics[height=4cm]{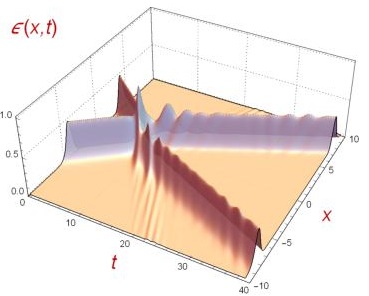}}
\caption{\small Energy density representation in the $K^{(q_1,q_2,-1)}$-$K^{(-q_1,-q_2,1)}$ scattering process for the initial velocities $v_0=0.2$ (left) and $v_0=0.4$ (right)}
\end{figure}

\noindent An intriguing point related to the previous description is the fate of the kink-antikink coupling: does it become a stable oscillatory state or do the kink lumps succumb to mutual annihilation, leaving a radiation vestige? With the purpose of investigating this question, the evolution of the total energy  in the simulation interval is plotted in Figure 10 together with the long term evolution  of the period of the motion. The numerical simulations indicate that the energy loss after a period is very small and decreases over time. The total energy for $t=5000$ is approximately $0.691125$. From a numerical point of view it is not possible to guarantee that this configuration is completely stable although we can affirm that the survival time of this state is long. Therefore we can infer that kink and antikink form a very lasting bound state. The evolution of period $T$ in this cyclic motion is shown in Figure 10. After a short interval of time the period $T$ tends to the value $2.77\pm 0.01$. This value seems to be independent of the shooting velocities $v_0$ in those cases where the bound state is formed. Notice however that the kink-antikink motion is not a pure oscillation because the frequency oscillates slightly. This behavior is pointed out in the framed plot inside Figure 10 (right) where a zoomed image of the period evolution curve is exhibited.

\begin{figure}[h]
\centerline{\includegraphics[height=2.5cm]{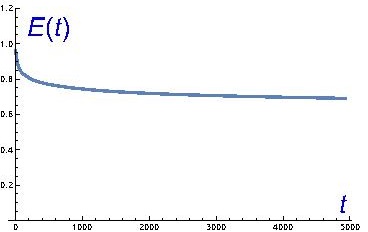} \hspace{1.5cm} \includegraphics[height=2.5cm]{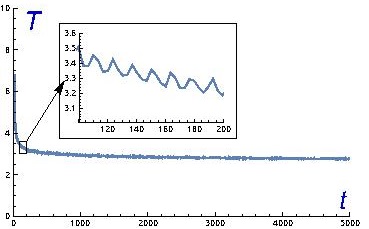}}
\caption{\small Evolution of the total energy of the $K^{(q_1,q_2,-1)}$-$K^{(-q_1,-q_2,1)}$ scattering process for the velocity $v_0=0.2$ in the simulation interval (left) and evolution of the kink-antikink motion period (right) in the time interval $[0,5000]$.}
\end{figure}

\item If the speed $v_0>v_c$ then the attraction force is not strong enough to attach the kink to the antikink. They collide and bounce back with an escape velocity $v_f$. This type of events has been illustrated in Figure 9 (right) for the case $v_0=0.4$. Notice that the kink-antikink collision is followed by radiation emission.
\end{itemize}

The global pattern is shown in the Figure 11 where the final velocity of the lumps is depicted with respect to its initial velocity. This graphic has been generated using a discrete number of points with step $\Delta v_0=0.001$. This study has been refined near the critical velocity $v_c$. Observe that if $v_0<v_c$ the final velocity is zero, which implies the formation of a bion (the kink and the antikink form a bound state). As before the dashed line describes an elastic scattering process.

\begin{figure}[h]
\centerline{\includegraphics[height=2.8cm]{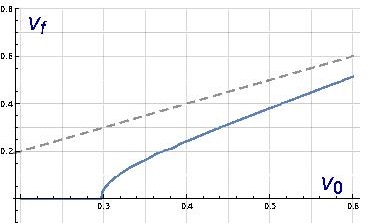}}
\caption{\small Graphical representation of the final velocity $v_f$ of the lumps as a function of the initial velocity $v_0$ in the $K^{(q_1,q_2,-1)}$-$K^{(-q_1,-q_2,1)}$ scattering.}
\end{figure}

Notice that in the kink-antikink scattering described in this Section the $N$-bounce reflection ($N\geq 2$) does not arise. A qualitative explanation of this fact is that the resonant energy transfer mechanism becomes effective when internal vibrational eigenmodes are present. We recall that in two coupled scalar field theories the kink fluctuation operator (\ref{hessiano}) is in general a $2\times 2$ non-diagonal matrix differential operator, whose spectrum must be usually identified numerically. As previously mentioned the basic kink fluctuation operator ${\cal H}[K_{\rm static}^{(q_1,q_2,\lambda)}(\overline{x})]$ lacks discrete eigenmodes other than the zero modes. The resonant energy transfer mechanism could also be activated by internal vibrational modes associated with the combined kink-antikink configuration, see \cite{Dorey2011:prl,Gani1999:pre,Weigel2014:jop}. In Figure 12 the potential well components of the kink-antikink fluctuation operator have been depicted. This operator comprises a continuous spectrum on the threshold value 2 in addition to two discrete eigenvalues which are approximately zero and come from the zero modes of the kink and the antikink. Therefore the collision between these kinks excites continuous eigenmodes which are responsible for the radiation phenomena. This heuristically justifies the absence of resonant windows in these scattering processes.

\begin{figure}[h]
\centerline{\includegraphics[height=2.8cm]{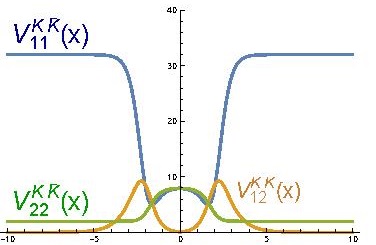}}
\caption{\small Potential well components $V_{ij}(x)$, $i,j=1,2$ of the second order small fluctuation operator associated with the combined kink-antikink configuration.}
\end{figure}

\subsection{The $K^{(q_1,q_2,1)}$-$K^{(-q_1,-q_2,-1)}$ scattering process}

Finally we shall describe the last class of two-lump scattering events. This process is similar to the previous one in the sense that we study the collision between a kink and an antikink of the same type although now they are arranged in the reverse order. Therefore the initial configuration consists of the concatenation of an antikink with charge $(q_1,q_2)$ followed in the $x$-axis by its own kink with charge $(-q_1,-q_2)$. A first difference with respect to the kink-antikink scattering explained in the previous subsection is that now the attractive forces are much weaker than in that event. This can be seen by the fact that the antikink-kink bound state arises when the collision velocity is much smaller than in the previous case. Indeed the critical velocity $v_c$ which divides the initial velocity regimes where the single lumps escape and where they are forced to collide a second time is given by
\[
v_c\approx 0.04162
\]
The general behavior of these scattering processes is illustrated in Figure 13 where the final velocity of the single lumps is represented as a function of the initial velocity $v_0$.

\begin{figure}[h]
\centerline{\includegraphics[height=6.cm]{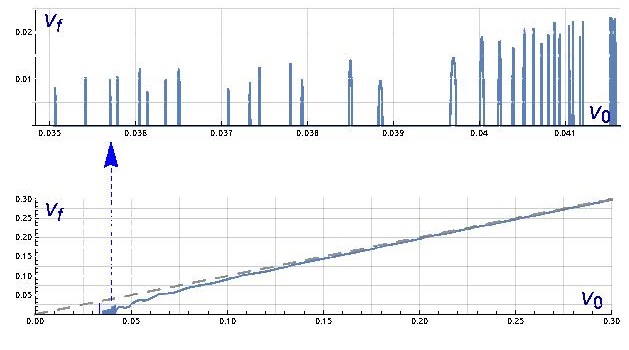}}
\caption{\small Graphical representation of the final velocity $v_f$ of the lumps as a function of the initial velocity $v_0$ in the $K^{(q_1,q_2,1)}$-$K^{(-q_1,-q_2,-1)}$ scattering. The top figure is a zoomed image of the small box remarked in the bottom figure where the 2-bounce resonant windows arise.}
\end{figure}

We can observe that:
\begin{itemize}
\item For the regime $v_0<v_c$ we find two possibilities. The first one is shown in Figure 14 (left) where the formation of a antikink-kink bion is depicted. In this simulation the basic lumps travel with a collision velocity $v_0=0.02$. After the first impact some vibrational modes are excited although radiation emission is negligible. Notice the small fluctuations in the energy density between the collisions. The single lumps do not manage to escape each other and remain trapped in a bound state. In the figure 13 this situation is characterized by a zero final velocity. The second possibility is described in the Figure 14 (middle) for the value $v_0=0.03970$. In this case the single lumps escape after the second collision. The energy accumulated in the internal vibrational mode is transferred back to the kinetic energy of the lumps, which breaks the bound estate. In this model this behavior arises for narrow initial velocity windows, which are called the 2-bounce resonance windows. In the Figure 13 (top) we plot the set of resonance windows which arise just below the critical velocity $v_c$. The existence of resonance windows has been explored using a step $\Delta v_0=0.00001$. The presence of other non-detected resonance windows, whose width is narrower than that value, is taken for granted due to the fractal nature of these windows sets, see \cite{Goodman2008:chaos, Goodman2005:sjoads, Goodman2007:prl}.

\begin{figure}[h]
\centerline{\includegraphics[height=3.8cm]{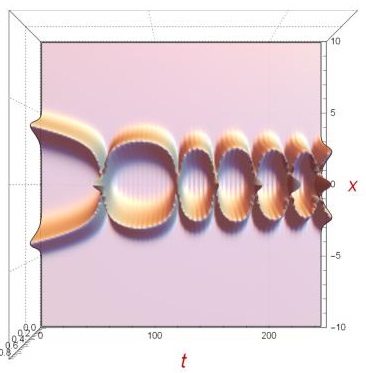} \hspace{0.2cm} \includegraphics[height=3.8cm]{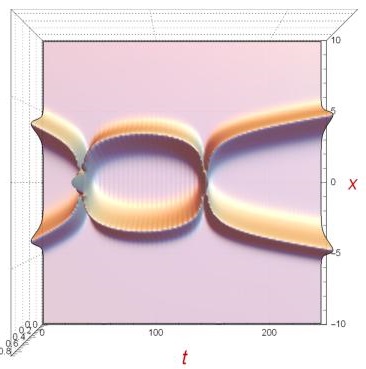} \hspace{0.2cm} \includegraphics[height=3.8cm]{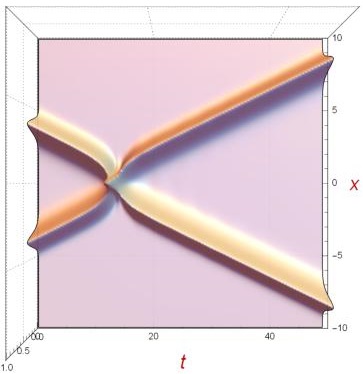}}
\caption{\small Energy density representation in the $K^{(q_1,q_2,1)}$-$K^{(-q_1,-q_2,-1)}$ scattering process for the initial velocities $v_0=0.02$ (left), $v_0=0.03970$ (middle) and $v_0=0.2$ (right).}
\end{figure}

\item For collision velocities $v_0>v_c$ the antikink-kink coupling is avoided. The simulation displayed in Figure 14 (right) for $v_0=0.2$ shows that the basic lumps attract each other, collide and bounce back. Finally the antikink and the kink move away. The global process is quite elastic even for high speeds $v_0$ because the radiation emission is small.

The explanation of the previous pattern can be understood by means of the resonant energy transfer mechanism although the internal vibrational eigenmode is associated with the combined antikink-kink configuration. In Figure 15 we have depicted the potential well components of the second order small fluctuation operator associated with a configuration following the form
\begin{eqnarray*}
\phi^{\overline{K}K}(x) &=& \phi_{\rm static}^{(q_1,q_2,1)}(x-x_1) + \phi_{\rm static}^{(-q_1,-q_2,-1)}(x-x_1) -{\textstyle\frac{1}{2}} \nonumber \\ \psi^{\overline{K}K}(x) &=& \psi_{\rm static}^{(q_1,q_2,1)}(x-x_1) + \psi_{\rm static}^{(-q_1,-q_2,-1)}(x-x_1) \label{akconf}
\end{eqnarray*}
which represents an antikink followed by a kink separated by a distance equals to $2x_1$. The second diagonal component $V_{22}^{\overline{K}K}(x)$ is a potential well whose asymptotic behavior tends to the value 8 while the valley floor level reaches the value 2. The well width increases as the antikink-kink distance increases. A continuous spectrum emerges on the threshold value $\omega^2=8$. In the Figure 15 (right) the discrete eigenvalues of the fluctuation operator associated to the previous configuration are plotted as a function of the magnitude $x_1$. We can observe that the number of discrete eigenmodes grows as the separation between the antikink and the kink increases. The potential well $V_{22}^{\overline{K}K}(x)$ becomes narrower as the lumps approach each other and only a few discrete eigenmodes survive this process. The resonant energy transfer mechanism can be activated by some of these excited modes making the resonant windows to arise.

\begin{figure}[h]
\centerline{\includegraphics[height=2.8cm]{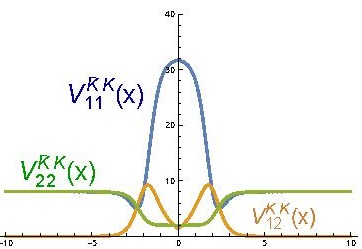} \hspace{1.5cm} \includegraphics[height=2.8cm]{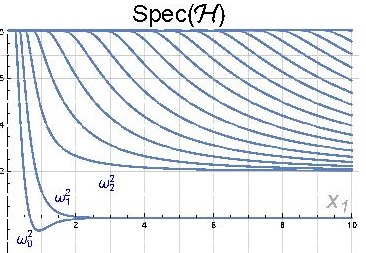}}
\caption{\small Potential well components $V_{ij}(x)$, $i,j=1,2$ (left) and discrete eigenvalues as a function of the lump separation (right) of the second order small fluctuation operator associated with the combined antikink-kink configuration (\ref{akconf}).}
\end{figure}
\end{itemize}

\section{Conclusions and further comments}

The study of the kink dynamics in a two coupled scalar field theory model in two space-time dimensions with potential term (\ref{potential}) has revealed a rich plethora of different interactions between the basic topological defects of the model. The static kink manifold unveils the existence of four basic particles together with its corresponding antiparticles described respectively by four single kinks and its antikinks. An important feature which rules the dynamics is the kink-antikink relation type. A kink and an antikink which live on the same trajectory are said to be of the type. As we have shown in the previous sections there exists an attractive force between this type of kinks. The kink ordering is also a property to be considered in these scattering processes. For instance, in the kink-antikink interaction there exists a critical velocity $v_c\approx 0.29703$ which distinguishes the initial velocity regimes where the bion formation and the lump reflection occur. In the antikink-kink interaction the velocity $v_c\approx 0.04162$ plays the same role but now two-bounce resonant windows arise just below this value. A qualitative explanation of these different behaviors underlies the small fluctuation operator spectrum valued on the combined kink-antikink or antikink-kink configurations. In the first case there is no internal vibrational eigenmodes and the continuous spectrum begins at the value 2. In the second case the continuous spectrum starts at the threshold value 8 and we can find vibrational modes in the range $[2,8]$. Therefore in the kink-antikink interaction the continuous eigenmodes are easily excited, which implies radiation emission while in the second case the vibrational modes play a predominant role involving the presence of the two-bounce resonant windows. In addition the kink-antikink interaction is stronger at short distances than the antikink-kink force although this last one has a longer range.

On the other hand repulsive forces manage the antikink-kink interaction when the involved lumps are of different type. Radiation emission is relevant for high speeds. In contrast the kink-antikink interaction is almost absent, its effect is only appreciable when the lumps are merged and compel the particles to concentrate its energy in a small region, such that the energy density peak quadruple the value of the single lump one. These two-body scattering processes conform the fundamental events in this model and they constitute the blocks which allow to explain more complex scattering processes.

As a final comment, it would be interesting to investigate the kink dynamics in other two scalar field theory models in order to acquire a more global perspective of the problem. The celebrated MSTB model \cite{Alonso1998:jopa} or its generalizations \cite{Alonso2008:pd} arise as natural candidates to this scrutiny. The study of the kink dynamics in massive nonlinear $S^2$-sigma models \cite{Haldane1983:prl} also constitutes a challenging problem.

\appendix

\section{Numerical analysis for kink scattering processes}

In this appendix we introduce the particular expressions obtained from the Strauss-Vazquez numerical scheme with Mur contour conditions adapted to study the evolution of a kink configuration ruled by the non-linear partial differential equations (\ref{euler01}) and (\ref{euler02}). Computational limitations compel us to restrict the space coordinate to the interval $[x_m,x_M]$ where we assume that the relevant kink scattering processes occur. The evolution of the phenomenon is studied in the time period $[0,T]$. We construct a finite mesh with $J$ space subintervals and $N$ time subintervals for the spacetime. With this notation the space and time steps are respectively given by
\[
\delta= \frac{x_M-x_m}{J} \hspace{0.5cm},\hspace{0.5cm} \tau=\frac{T}{N}\hspace{0.3cm}.
\]
We shall denote $\phi_j^{n}=\phi(x_m+j\,\delta,n \tau)$ and $\psi_j^{n}= \psi(x_m+j\,\delta,n \tau)$, the values of the fields at the mesh points. We recall that the total energy (\ref{totalenergy}) is an invariant magnitude for the scalar field theories which we are dealing with. This fact suggests the use of the energy conservative implicit second order numerical scheme
\begin{eqnarray}
\frac{\phi_j^{n+1} - 2 \phi_j^n + \phi_j^{n-1}}{\tau^2} - \frac{\phi_{j+1}^{n} - 2 \phi_j^n + \phi_{j-1}^n}{\delta^2} + \frac{U[\phi_j^{n+1},\psi_j^n]-U[\phi_j^{n-1},\psi_j^n]}{\phi_j^{n+1} - \phi_j^{n-1}} =0, && \label{algo01} \\
\frac{\psi_j^{n+1} - 2 \psi_j^n + \psi_j^{n-1}}{\tau^2} - \frac{\psi_{j+1}^{n} - 2 \psi_j^n + \psi_{j-1}^n}{\delta^2} + \frac{U[\phi_j^n,\psi_j^{n+1}]-U[\phi_j^n,\psi_j^{n-1}]}{\psi_j^{n+1} - \psi_j^{n-1}} =0, && \label{algo02}
\end{eqnarray}
where $U$ stands for the potential term $U(\phi,\psi)$ given in (\ref{potential}). This is the adaptation of the Strauss-Vazquez scheme introduced in \cite{Strauss1978:jocp,Jimenez1990:amac,PenYu1983:jas} to our two-scalar field theory context. By construction this algorithm preserves the discrete total energy
\begin{eqnarray}
E^n&=&\sum_{j} \delta \Big[ \frac{1}{2\tau^2} (\phi_j^{n+1}-\phi_j^n)^2 + \frac{1}{2\delta^2} (\phi_{j+1}^{n+1} -\phi_j^{n+1})(\phi_{j+1}^{n} -\phi_j^{n}) + \nonumber \\ && \hspace{0.5cm} + \frac{1}{2\tau^2} (\psi_j^{n+1}-\psi_j^n)^2 +  \frac{1}{2\delta^2} (\psi_{j+1}^{n+1} -\psi_j^{n+1})(\psi_{j+1}^{n} -\psi_j^{n}) + \label{discreteenergy} \\ && \hspace{0.5cm} +\frac{1}{2} \Big(U[\phi_j^{n+1},\psi_j^n]- U[\phi_j^n,\psi_j^{n+1}] \Big) \Big]\nonumber
\end{eqnarray}
which can be understood as a discretization of the total energy (\ref{totalenergy}). One reason for the convenience of the numerical method introduced in (\ref{algo01}) and (\ref{algo02}) underlies the fact that some kink scattering processes involve radiation phenomena where linear plane waves are emitted and travel with large speeds in both spatial directions. The use of an energy conservative numerical method allows us to control the amount of energy which escapes through the frontiers of our finite interval $[x_m,x_M]$, the only possibility of energy change in our numerical scheme. This strategy must be complemented with the use of absorbing contour conditions. If the spatial interval is large enough to the kink scattering occurs far away from the boundaries the dynamics in these peripheral regions is described by the linear partial differential equations
\[
\frac{\partial^2 \phi}{\partial t^2} - \frac{\partial^2 \phi}{\partial x^2} =  0 \hspace{1cm}, \hspace{1cm}
\frac{\partial^2 \psi}{\partial t^2} - \frac{\partial^2 \psi}{\partial x^2} = 0\hspace{0.3cm}.
\]
This fact suggests the use of second order absorbing Mur contour conditions for our problem \cite{Mur1981:emc}, which are given by the relations
\begin{eqnarray}
&& \phi_0^{n+1}-\phi_1^n - \frac{n_c-1}{n_c+1} (\phi_1^{n+1}-\phi_0^n)=0 \label{mur1} \hspace{0.3cm},\\
&& \phi_J^{n+1}-\phi_{J-1}^n - \frac{n_c-1}{n_c+1} (\phi_{J-1}^{n+1} - \phi_J^n) =0 \label{mur2} \hspace{0.3cm}, \\
&& \psi_0^{n+1}-\psi_1^n - \frac{n_c-1}{n_c+1} (\psi_1^{n+1}-\psi_0^n)=0 \label{mur3}\hspace{0.3cm}, \\
&& \psi_J^{n+1}-\psi_{J-1}^n - \frac{n_c-1}{n_c+1} (\psi_{J-1}^{n+1} - \psi_J^n) =0 \label{mur4}\hspace{0.3cm},
\end{eqnarray}
where $n_c=\tau/\delta$. These contour conditions have the ability of absorbing the radiation which arrives to the frontiers of our simulations. The initial conditions
\[
\phi(x,0)=f_0(x) \hspace{0.2cm},\hspace{0.2cm} \psi(x,0)=g_0(x)\hspace{0.2cm}, \hspace{0.2cm} \frac{\partial \phi}{\partial t}(x,0)=f_1(x)\hspace{0.2cm} ,\hspace{0.2cm} \frac{\partial \psi}{\partial t}(x,0)=g_1(x)
\]
let start the algorithm by fixing
\begin{eqnarray*}
\phi_j^0 &=& f_0(x_m+j \,\delta) \hspace{0.3cm},\\
\psi_j^0 &=& g_0(x_m+j \, \delta) \hspace{0.3cm},\\
\phi_j^1 &=& \phi_j^0 + \tau \, f_1(x_m+j\,\delta)+\frac{1}{2} \Big(\frac{\tau}{\delta} \Big)^2 [\phi_{j-1}^0 -2 \phi_j^0 + \phi_{j+1}^0] -\frac{\tau^2}{2}\, \frac{\partial U}{\partial \phi}(\phi_j^0,\psi_j^0)\hspace{0.3cm}, \\
\psi_j^1 &=& \psi_j^0 + \tau \, g_1(x_m+j\,\delta)+\frac{1}{2} \Big(\frac{\tau}{\delta} \Big)^2 [\psi_{j-1}^0 -2 \psi_j^0 + \psi_{j+1}^0] -\frac{\tau^2}{2}\, \frac{\partial U}{\partial \psi}(\phi_j^0,\psi_j^0)\hspace{0.3cm},
\end{eqnarray*}
which corresponds to a second order approximation consistent with the numerical scheme (\ref{algo01}) and (\ref{algo02}).

Finally we have to implement a procedure for estimating the error derived from the numerical method. This error control will be accomplished firstly by monitoring the evolution of the model invariants. The algorithm has been constructed to keep the total energy $E(t)$ constant but the total momentum $M(t)$ given by (\ref{totalmomentum}) should also be a constant of motion. In this sense we shall supervise the evolution of the discrete version of the total momentum
\begin{equation}
M_n=\sum_{\ell}  \frac{1}{2\tau} \Big[(\phi^{n+1}_j-\phi^{n}_j) (\phi^{n+1}_{j+1} - \phi^{n+1}_{j-1}) +  (\psi^{n+1}_j-\psi^{n}_j) (\psi^{n+1}_{j+1} - \psi^{n+1}_{j-1})  \Big]
\label{discretemomentum}\hspace{0.3cm}.
\end{equation}
A significant variation of the magnitude (\ref{discretemomentum}) along the time would indicate that the algorithm fictitiously accelerates the particles. In addition to this protocol we shall also analyze the difference between the results obtained by two simulations where the second one halves both the space and time steps used in the first one. In this sense we construct the functions
\begin{eqnarray}
\xi_1(t)&=&\max_{j=0,\dots,L} \left\{ \left| \phi^{n}_j (\delta,\tau) - \phi^{2n}_{2j} ({\textstyle \frac{1}{2}}\delta, \textstyle{\frac{1}{2}} \tau)\right| \right\} \nonumber \\
\xi_2(t)&=&\max_{j=0,\dots,L} \left\{ \left| \psi^{n}_j (\delta,\tau) - \psi^{2n}_{2j} ({\textstyle \frac{1}{2}}\delta, \textstyle{\frac{1}{2}} \tau)\right| \right\} \label{xi}
\end{eqnarray}
which give the maximum discrepancy between the values of the field components on the set of all the mesh points for every instant $t$. In (\ref{xi}) the notation $\phi^{n}_j (\delta,\tau)$ stands for the value of the field obtained by the simulation with space and time steps $\delta$ and $\tau$ respectively. A standard choice of the parameters in our simulations is given by the values $x_m=-10$, $x_M=10$, $T=50$, $J=16000$ and $N=160000$. This involves a value of the steps $\delta=0.00125$ and $\tau=0.003125$. A upper bound for the error parameters $|\xi_i(t)|$ for this setting is $5\times 10^{-5}$, that is, $|\xi_i(t)|\leq 5\times 10^{-5}$ for most of the simulations. The error estimation of the numerical procedure is measured by the difference between the results of two simulations where the space and time steps have been halved. The evolution of the total momentum is also monitored. This magnitude must be constant over time because it is an invariant of the system.

\section*{Acknowledgments}

The authors acknowledge the Spanish Ministerio de Econom\'{\i}a y Competitividad for financial support under grant MTM2014-57129-C2-1-P. They are also grateful to the Junta de Castilla y Le\'on for financial help under grant VA057U16.


\end{document}